\documentstyle[11pt]{article}

\newfont{\msbm}{msbm10}
\def\g{\gamma}

\def\U{{\cal U}}

\def\A{{\cal A}}

\def\N{\hbox{\msbm N}}

\def\Z{\hbox{\msbm Z}}

\def\R{\hbox{\msbm R}}

\def\C{\hbox{\msbm C}}

\def\H{{\cal H}}

\def\m{\mu}

\def\vf{\varphi}

\def\l{\lambda}

\newtheorem{prop}{Proposition}
\newtheorem{cor}{Corolary}
\newtheorem{theo}{Theorem}
\newtheorem{deff}{Definition}

\def\bc{\begin{cor}}
\def\ec{\end{cor}}

\def\bt{\begin{theo}}
\def\et{\end{theo}}

\def\bd{\begin{deff}}
\def\ed{\end{deff}}

\def\bp{\begin{prop}}
\def\ep{\end{prop}}

\def\ba{\begin{eqnarray}}
\def\ea{\end{eqnarray}}

\def\be{\begin{equation}}
\def\ee{\end{equation}}

\newfont{\msbms}{msbm6}  
\def\ab{\mbox{$\bar \A$}}

\def\br{\bar{\R}}
\def\bx{\mbox{$\bar x$}}

\def\rep{representation}
\def\reps{representations}

\begin{document}

%%%%%%%%%%%%%%%%%%%%%%%%%%%%%%%%%%%%%%%%%%%%%%%%%%%%%%%%%%%%%%%%%%%%%%
\title{The quantum configuration space of loop quantum cosmology}

\author{J M Velhinho}

\date{{ Departamento de F\'\i sica, Universidade da Beira 
Interior\\R Marqu\^es D'\'Avila e Bolama,
6201-001 Covilh\~a, Portugal}\\{jvelhi@ubi.pt}}

\maketitle
%%%%%%%%%%%%%%%%%%%%%%%%%%%%%%%%%%%%%%%%%%%%%%%%%%%%%%%%%%%%%%%%%5

%%%%%%%%%%%%%%%%%%%%%%%%%%%%%%%%%%%%%%%%%%%%%
\begin{abstract}

\noindent  
The 
article 
gives
an account of several aspects of the space known
as the Bohr compactification
of the line, featuring as the quantum configuration space in loop quantum cosmology,
as well as of the corresponding configuration space realization of the so-called  
polymer representation. 
Analogies with loop quantum gravity are explored,
providing an introduction to (part of)
the mathematical structure of loop quantum gravity, in a technically simpler context.
\end{abstract}
%%%%%%%%%%%%%%%%%%%%%%%%%%%%%%%%%%%%%%%%%%%%%%
%%%%%%%%%%%%%%%%%%%%%%%%%%%%%%%%%%%%%%%%%%%%

%%%%%%%%%%%%%%%%%%%%% BEGIN PAPER %%%%%%%%%%%%%%%%%%%%%%%%%

%%%%%%%%%%%%%%%
%%%%%%%%%%%%%%%%%%%%%%%%%%%%%%%%%%%%%%%%%
\section{Introduction}
\label{intro}
Loop quantum cosmology (LQC) is a canonical approach
to the quantization of symmetry reduced, typically homogeneous, gravitational
models. LQC
originated from the works of Bojowald (see~\cite{B} and references therein),
was further developed by Ashtekar, Bojowald and Lewandowski~\cite{ABL},
and more recently by Ashtekar, Pawlowski and Singh~\cite{APS}. 
(See also~\cite{BCK}  for new developments.)

LQC is 
inspired by loop quantum gravity\footnote{For reviews of loop quantum gravity see~\cite{AL,T,T2}.} 
(LQG), importing (whenever possible) ideas, methods
and structures used in LQG to the context of quantum cosmology.
In particular, in the quantization of the gravitational sector of homogeneous models,
LQC uses  a \rep\ of the (standard finite dimensional) Weyl algebra which fails to produce 
a continuous \rep\ of the configuration part of the Heisenberg-Weyl group.\footnote{Representations
of Weyl relations violating continuity conditions are sometimes called {\it non-regular}
in the mathematical-physics literature. Discussions and applications of such \reps, both
for infinite and finite dimensional systems, can be found e.g.\ in~\cite{AMS,CMS}.}
The LQC quantization is therefore non-equivalent to the usual Schr\"{o}dinger quantization.

The \rep\ of the Weyl relations used in LQC - called in this context the polymer representation -
is typically presented in its momentum space version, with momentum operators such as the triads
acting diagonally, and connection related configuration variables acting as translation operators.

On the other hand, it is known that
a configuration space realization of the polymer
\rep\ exists. However, even though only a finite number of degrees of freedom is involved, the polymer
\rep\ requires an extension of the classical configuration space, in order that configuration operators
can be realized as functions, or multiplication operators. 
Restricting attention to isotropic models, so that the classical configuration space
is the real line $\R$, the role of such a ``quantum 
configuration space'' is in this case played by the so-called Bohr compactification of the 
line.

The Bohr compactification of the line, which we will denote by $\br$,  
is a well known space within the mathematics community (see e.g.~\cite{R}), with several equivalent characterizations.
Not surprisingly, 
some of these characterizations are analogous to those of the space of generalized connections $\ab$,
featuring  as the 
(kinematical) 
quantum configuration space in LQG.\footnote{For reviews giving a detailed account of the space
$\ab$ see e.g.~\cite{T,V}.}

The present article provides a unified
account of several  aspects of the quantum configuration space $\br$, 
together with an explicit construction of the configuration space version of the 
polymer \rep.\footnote{Descriptions of the configuration version of the polymer representation
more detailed than previous ones in the literature appeared recently~\cite{A,CVZ2}.}  
Moreover, we explore  the analogies with the space $\ab$, aiming at an introduction to (part of) the mathematical structure of LQG,
in a technically simpler context.

Necessarily, the article  revisits  constructions and arguments previously put forward in the context of LQG.
On the other hand, 
%though retaining the same mathematical flavour, 
we will obviate 
detailed technical proofs, putting the emphasis  on displaying typical structures
instead.
We will focus exclusively on kinematical aspects and corresponding structures, leaving 
aside the hamiltonian constraint operator and the final picture provided by it.

The  article is organized as follows. Section \ref{poly} reviews the basics of
the polymer \rep\ used in LQC 
%in the quantization of the gravitational sector 
%of homogeneous and isotropic models 
(see~\cite{ABL,AFW,ALS,Vlqc,CVZ2} for more details and discussion). A 
%brief 
review of the 
%$C^*$-algebraic origins 
emergence 
of the Bohr compactification 
%$\br$ 
and its natural structures is also given.
Section \ref{group1} describes $\br$ as a space of (homo)morphisms. Up to some point, 
this corresponds to 
%the functorial
a similar
characterization of $\ab$. However, $\br$ is a
%a well known compact 
group, so it is
equipped with more structure, and in particular with the Haar measure.
%\footnote{This description thus corresponds more closely to the space of generalized $U(1)$ connections.}
The configuration space realization of the polymer \rep\ is then presented in section \ref{group2}.
In section  \ref{pro} we discuss the projective characterization of $\br$,
% as a limit of a family of tori, 
directly related to 
%a similar  
an important
structure
in LQG. In section \ref{spect} the space $\br$ is seen as the
spectrum of a classical configuration algebra, corresponding to the original construction
of $\ab$. 
Finally, one appendix is included, where the usual Schr\"{o}dinger \rep\ is presented in this context,
by means of  a different,
non-equivalent measure on $\br$.

\section{Polymer representation}
\label{poly}
\subsection{Momentum space formulation}
\label{poly2}
Let us consider the phase space $T^*\R\equiv \R^2$ of a particle in the line, coordinatized 
by a configuration variable $x$ and a canonical momentum variable $p$.
The so-called polymer representation of the associated Weyl relations is defined as follows,
using a momentum space formulation.
Let $\H_{\cal P}$ be the non-separable
Hilbert space spanned by mutually orthogonal vectors $|p\rangle$, $p\in\R$,
$\langle p'|p\rangle=\delta_{p' p}$, where $\delta_{p' p}$ is the Kronecker delta.
A general element of $\H_{\cal P}$ is of the form 
\be
\label{1}
\sum_{p\in\R}\psi(p)|p\rangle\ \ \rm{with} \ \ \sum_{p\in\R}|\psi(p)|^2<\infty.
\ee
%%%%
Thus, the polymer Hilbert space $\H_{\cal P}$ can also be described as the space of complex functions on $\R$ that are square integrable
with respect to the discrete measure. Necessarily, any wave function $\psi(p)$ can be non-zero only
on a countable subset of $\R$. 

The momentum operator $\hat p$ is defined in this \rep\ by:
\be
\label{3}
\hat p |p\rangle=p |p\rangle \ \ \ \ {\rm or}\ \ \ \ \hat p\psi(p)=p\psi(p).
\ee
Since the discrete measure is translation invariant, there are also well defined unitary operators 
$\U(k)$ implementing translations in momentum space:
\be
\label{2}
\U(k)|p\rangle=|p+k\rangle\ \ \ \ {\rm or}\ \ \ \ \U(k)\psi(p)=\psi(p-k),\ \ k\in \R.
\ee

The exponentiation of the operator $\hat p$  together with the operators $\U(k)$ provide a representation of
the  Weyl relations, which is moreover irreducible. However,  the \rep\  of $\R$ given by $k\mapsto \U(k)$ is clearly not continuous.
In fact, for arbitrarily small $k$, a vector $|p\rangle$ is mapped by $\U(k)$ to an orthogonal one
$|p+k\rangle$. Thus, the  generator which would
correspond to the configuration operator $\hat x$ is not defined. 

The operators $\U(k)$ can nevertheless be seen as giving a 
quantization of the classical configuration functions $e^{ikx}$. Reality conditions are satisfied, since 
$\U^{\dag}(k)=\U(-k)$. Thus, the polymer \rep\ provides
a quantization, in the usual Dirac sense, of the Poisson algebra of phase space functions
made of finite linear combinations of the functions $p$ and $e^{ikx}$, $k\in\R$.
In particular, the configuration part of the Poisson algebra is the linear space of 
continuous and bounded complex functions in $\R$ of
the form
%%%%%%%%%%
\be
\label{new23}
\R\ni x\mapsto f(x)=\sum_j c_j e^{ik_jx},
\ee
where the sums are finite, $k_j$ are arbitrary real numbers and $c_j$ are complex coefficients.
The set of functions (\ref{new23}) clearly separates points in $\R$, i.e.~given $x$, $x'\in\R$, $x\not = x'$,
one can find a function $f$ such that $f(x)\not = f(x')$. In fact, two functions are sufficient to separate
points, e.g.~$e^{ik_1x}$ and $e^{ik_2x}$, with $k_1/k_2$ being an irrational number.

In this representation, the spectrum of the momentum operator $\hat p$
is the real line,  but  each point of $\R$ belongs to the discrete spectrum,
rather than to the continuous spectrum, since well defined eigenvectors $|p\rangle$
exist. In this sense, the ``quantum momentum space'' in the polymer representation can
be seen as $\R$ equipped with the discrete topology. 

Note that by switching the roles of
configuration and momentum one obtains a non-equivalent \rep, where the Bohr compactification 
corresponds to momentum space. 
With the variable $x$ corresponding to connections
and $p$ to triads, the polymer \rep\ used in homogeneous and isotropic 
LQC~\cite{ABL} is the one described above.

Further regarding the correspondence of the present case to LQG, the functions $x\to e^{ikx}$
are playing the role of holonomies, indexed by real numbers $k$, which then correspond to
edges.

\subsection{From momentum space to configuration space: an overview}
\label{motiv}
Before we start displaying the structure of the Bohr compactification of the line, let us first review
its emergence 
%of the space and its natural structures 
from the  perspective of $C^*$-algebras, 
which is the approach followed in  loop quantization. A brief overview of the remaining
part of 
the article is also given.

The configuration version of the polymer \rep\ is, of course, a unitarily equivalent \rep\
such that the configuration variables  (\ref{new23}) are realized as multiplication operators.
Note first that the linear space of variables  (\ref{new23}) is actually a commutative
$*$-algebra with identity, with respect to multiplication of functions and complex conjugation. 
This algebra, let us call it  $\cal C$, is obviously represented by bounded operators in $\H_{\cal P}$.
%by $e^{ikx}\mapsto\U(k)$. 
Moreover, the \rep\ is cyclic, i.e.~there is a vector, e.g.~$|0\rangle\in\H_{\cal P}$,
such that the action of (the \rep\ of)
$\cal C$ on $|0\rangle$ produces a dense set.

One can now take advantage of the general theory of commutative $C^*$-algebras.
In a generic situation, e.g.~when looking for \reps, one would need to 
%faithfully 
complete our $*$-algebra into a $C^*$-algebra,
by means of an appropriate norm. In the present case we already have the (equivalence class of the) \rep,
so the obvious thing to do is to consider the operator norm in the polymer \rep, and the corresponding $C^*$-algebra
of bounded operators generated by the unitaries $\U(k)$.
Not surprisingly, this $C^*$-algebra  coincides (up to isomorphism) with the 
Bohr algebra of almost periodic functions, i.e.~the algebra of (bounded and continuous) functions in $\R$
obtained by completing $\cal C$
with respect to the supremum norm. Let us denote this $C^*$-algebra by $\bar {\cal C}$,
%(Also the same $C^*$-algebra is obtained if we use the operator
%norm in the Schr\"{o}dinger \rep.)
corresponding to the holonomy algebra in LQG.

General  results now tell us that any commutative $C^*$-algebra (with identity) $A$ can be seen as the algebra 
$C(\Delta)$ of all continuous
functions on a compact space $\Delta$, called the spectrum of the algebra. The isomorphism that maps
elements of $A$ to functions in the spectrum is called the Gelfand transform. When, as in the present case,
the algebra $A$ is an algebra of functions in a space $S$, and the algebra separates points, the Gelfand transform also gives an 
injective map $S\to \Delta$, whose image is dense. The Bohr compactification
can be seen as a particular case of this Gelfand compactification. In the present case, the spectrum of 
the algebra $\bar {\cal C}$ is the
space $\br$, the Bohr compactification of the line, which then contains  $\R$ as
a dense subset.

Finally, any cyclic \rep\ of the algebra
of continuous functions $C(X)$ on a compact space $X$ can be naturally realized in a Hilbert
space $L^2(X,\m)$ of square integrable functions, for some normalized measure $\m$ on $X$.

Putting it all together, it is guaranteed to exist a measure $\m_0$ on $\br$ such that the polymer
\rep\ of  $\bar {\cal C}$ is unitarily equivalent to the \rep\ in $L^2(\br,\m_0)$, with elements of
$\bar {\cal C}$ acting as multiplication operators.

Note that the knowledge of the spectrum of the configuration algebra alone does not tell us much
about the actual ``quantum configuration space'' for a given \rep. This is determined by the support
of the corresponding measure.  In the case of the polymer \rep\ the extension from
the classical space $\R$ to $\br$ is in fact essential.
This situation is typical of quantum field
theories, including LQG. On the other hand,  one can  check that the measure
in $\br$ corresponding to the Schr\"{o}dinger \rep\ is supported on the classical configuration
space $\R$.

\bigskip

As in LQG,
a projective description of $\br$ can 
%also 
be given.
%, probably the most pictorial one. 
This is  related to the inductive structure of the algebra $\cal C$ (\ref{new23}).
This structure is provided by the family of the finitely generated $*$-subalgebras, whose union
is the whole algebra $\cal C$. In a natural way, the spectrum of (the completion of)
each of these algebras is a torus, its dimension being equal to the number of generators.
As we consider larger and larger subalgebras, one can see the spectrum $\br$ as the limit
of a family of tori, of growing dimensions. 
(Note, however, that projective-inductive techniques do not play in LQC the same prominent role
as in LQG.)

Again on the algebraic side, a natural characterization of $\br$ is related to the properties
of the basic elements of $\cal C$, namely the exponential functions. These are labelled by
real numbers $k$, and map sums into products. Thus, exponentials define  (homo)morphisms
from $\R$ to the unit circle in $\C$, which are in particular continuous. It turns out 
that $\br$ coincides with the set of all, not necessarily continuous,
homomorphic maps from $\R$ to the unit circle.

This characterization of $\br$ is  well known in harmonic analysis, 
where it is seen as the dual group of the discrete group $\R$ (see e.g.~\cite{R}).
The relation with the $C^*$-algebra approach is that  $\bar {\cal C}$ is essentially generated by this
discrete group.
Thus, in particular, $\br$ is a commutative compact group, and comes equipped with the Haar measure,
which is precisely the measure $\m_0$ corresponding to the polymer \rep.
From this perspective, the unitary transformation from $\H_{\cal P}$ to  $L^2(\br,\m_0)$ is a 
Fourier transformation, and it is actually quite easy to perform. 

Although this group structure is not present in the space of generalized connections $\ab$ 
in LQG,\footnote{It is present in the case of generalized $U(1)$ connections~\cite{AL1,M,Va,Vc,ALfock}, which is then much closer to the present case.}
its characterization as  a set of (homo)morphisms  is, in fact, one of the three
known characterizations. So, this description too has an analogue in LQG, and we actually start from it,
as its early introduction facilitates further discussion. 
  
Note also that introducing the $\m_0$ measure by projective-inductive techniques 
(corresponding to the original introduction of the Ashtekar-Lewandowski measure in LQG), although possible, 
is not natural here, and we therefore start from the given Haar measure in $\br$. On the other hand, we do explicitely confirm
that the Gelfand spectrum of $\bar {\cal C}$ coincides with the dual of the discrete group $\R$.
In particular concerning finitely generated subalgebras (which is the essence of the argument), this discussion
provides an easy to follow example of the Gelfand compactification, in finite dimensions.   

%%%%%%%%%%%%%%%%%%%%%%%%%%%%%
%%%%%%%%%%%%%%%%%%%%%%%%%%%%%%%%
\section{Configuration space realization}
\label{group}
The space $\br$ is introduced as a set of (homo)morphisms, corresponding to 
a similar characterization
of $\ab$ (see~\cite{AL1,AL2,AL3,Ba} for the origins and~\cite{Vg,V} for reformulation and review). The role of the group of hoops 
(or the groupoid of paths) is here played
by the discrete group $\R$. The group $SU(2)$ is  replaced by $T$, the unit circle in the complex plane $\C$. 

As mentioned, the present characterization of $\br$ is precisely that of a dual group. 
Although we do not insist
on this perspective, 
the group structure
of $\br$ 
is introduced from the start. Furthermore, basic results in harmonic analysis will necessarily emerge,
in particular in section \ref{group2},
where
the configuration version of the polymer \rep\ is presented. 
In the following, we will refer to homomorphisms simply as morphisms.

\subsection{Quantum configuration space as a compact group}
\label{group1}
Let us consider the real line $\R$  equipped with the commutative group structure
given by addition of real numbers. The Bohr compactification $\br$ can be described as the set 
${\rm Hom}[\R,T]$ of all, not necessarily continuous, group
morphisms from the  group $\R$ to the multiplicative  group $T$ 
of unitaries in $\C$. Since we are 
going to introduce the topology of
$\br$ immediately after, we identify it with ${\rm Hom}[\R,T]$  right from the start:
%%%%%%%%%%%%%%%
\be
\label{4}
\br\equiv{\rm Hom}[\R,T].
\ee
%%%%%%%%
The generic element of $\br$ will be denoted by \bx. So, every $\bx\in\br$ is a map, 
$\bx:\R\to T$ such that
%%%%%%%%
\be
\label{5}
\bx(0)=1\ \ \ {\rm and}\ \ \ \bx(k_1+k_2)=\bx(k_1)\bx(k_2), \ \forall k_1,k_2\in\R.
\ee
%%%%%%%%%%%%

Since $T$ is a commutative group, it is clear that 
%%%%%%
\be
\label{6}
\bx\, \bx'(k):=\bx(k)\bx'(k)
\ee
%%%%%%%%%%%%
defines a group structure on $\br$, which is moreover commutative.
The group $\br$ is a subgroup of the group of all (not necessarily morphisms) maps from
$\R$ to $T$. Since this group of maps can be identified with the product group
$\times_{k\in\R}T$, and $T$ is compact, it carries  the Tychonoff product topology,
with respect to which $\times_{k\in\R}T$ becomes a compact 
(Hausdorff)\footnote{All spaces considered are Hausdorff spaces, so we will refrain from mentioning it.} 
group.
This structure descends to the subgroup $\br$, making it a topological group.
Moreover, from the fact that $\br$
contains only morphisms, one can see that it is a closed subset  of $\times_{k\in\R}T$,
and it is therefore compact.
 
Thus, $\br$ is a commutative compact  group, with respect to the group operation (\ref{6})
and the Tychonoff topology. A more explicit description of the topology is the following. 
For each $k\in\R$, let us consider the function $F_k:\br\to T$
%\subset\C 
defined by
%%%%%%%%%
\be
\label{7} 
F_k(\bx)=\bx(k).
\ee
The functions (\ref{7}) are continuous and, in fact,  the Tychonoff
topology in $\br$ is precisely the weakest topology such that all functions (\ref{7}),
$\forall k\in\R$, are continuous.

As in the case of connections in LQG, there is a natural map from the classical configuration space
to the quantum configuration space, let us call it $\Theta$, defined as follows (we will confirm in section~\ref{spect}
that $\Theta$ is indeed the Gelfand - Bohr compactification):
%%%%%%
\be
\label{theta}
\Theta :\R\to \br,\ \ y\mapsto {\bar x}_y,\   \ {\bar x}_y(k):=e^{iyk}\ \ \forall k\in\R.
\ee
Since this map
is  injective, the set $\br$ can
be seen as an extension of $\R$. The classical configuration space then appears as a subspace,
of continuous morphisms, of the space of all morphisms $\br$. In addition, the image $\Theta(\R)$ is a dense subset. It is worth
mentioning that the injection $\Theta$ is continuous, but is not an homeomorphism into
its image. The topology induced on $\R$ as a subset of $\br$, which can be seen as the weakest topology
such that all functions $x\mapsto e^{ikx}$, $k\in \R$, are continuous, is weaker than the usual topology.

Note that, since we know in advance that $\br$ is the spectrum of the $C^*$-algebra $\bar {\cal C}$, it is guaranteed
that  any almost periodic function in $\R$ can be extended from the dense set $\Theta(\R)$ to a continuous function in $\br$ 
(this is the Gelfand transform in this case). In particular, the functions
$F_k$ correspond to the exponential functions $\R\ni x\mapsto e^{ikx}$. However, especially upon introduction of the measure
defining the \rep, this correspondence is to be seen in the sense of operators, i.e.~it is when 
considered as a multiplication
operator that $F_k$ corresponds to the classical function $e^{ikx}$, giving in fact its quantization.

\subsection{Polymer representation in configuration space}
\label{group2}
Being a compact group, $\br$ is equipped with a normalized invariant (under the group operation) measure,
namely the Haar measure, which we denote by $\m_0$.
The Haar measure gives precisely the configuration space realization of the polymer \rep, as we now
show.

Let us consider the Hilbert space 
of square integrable functions 
$L^2(\br,\m_0)$, which we will denote by $\H_0$.
Since the functions $F_k$ (\ref{7}) are continuous, they are in particular integrable. As expected, they form 
a complete orthonormal set in $\H_0$. One can easily see from the invariance of the measure that the set is orthonormal.
In fact, for every fixed $\bar x'\in\br$ we have:
%%%
\be
\label{z1}
\int_{\br} F_k(\bar x'\bar x) d\m_0(\bar x)=\int_{\br} F_k(\bar x) d\m_0(\bar x),
\ee
leading to
%%%%
\be
\label{z2}
\left(1-\bar x'(k)\right) \int_{\br} F_k d\m_0=0.
\ee
Since this is true $\forall \bar x'\in\br$, we conclude that
%%%%%%%%%%%
\be
\label{8}
\int_{\br} F_k d\m_0 = \delta_{k 0}, \ \ \ k\in\R,
\ee
where $\delta_{k 0}$ is the Kronecker delta. 
From $F_k^*=F_{-k}$ and $F_kF_{k'}=F_{k+k'}$ 
follows that 
$\{F_k,\ k \in\R\}$ is an orthonormal set. 
It is also straightforward to confirm that this set is complete,
since the space of finite linear combinations of functions $F_k$ is a $*$-subalgebra of the algebra
of all continuous functions $C(\br)$, contains the identity function, and  separates points in $\br$.
The Stone-Weierstrass theorem then ensures that this linear space is dense in $C(\br)$, with respect to the supremum norm.
(In alternative, this linear space corresponds to $\cal C$, which is dense in $\bar {\cal C}$ by construction.)
Standard arguments show that it is necessarily dense with respect to the 
$L^2$-norm.\footnote{For compact $X$, $C(X)$ is $L^2$-dense for any regular Borel measure.
On the other hand, uniform convergence implies $L^2$-convergence.}

Thus, the Hilbert space $\H_0$ is isomorphic to the polymer space $\H_{\cal P}$, the 
unitary transformation ${\cal T}:\H_{\cal P}\to \H_0$ being  given by the map between bases:
%%%%%%%%%%
\be
\label{9}
{\cal T}\, :\, |p\rangle\mapsto F_p,\ \ \ \forall p\in\R.
\ee

The Hilbert space $\H_0$ provides a \rep\ of the $C^*$-algebra  $C(\br)$ which is faithful, since
the invariant Haar measure is faithful (i.e.~every non-empty open set has non-zero measure).
Denoting in particular the representative of $F_k\in C(\br)$  by $\Pi(k)$ we therefore have:
%%%%%%%%%%%%
\be
\label{10}
\Pi(k)\psi(\bx)= F_k(\bx)\psi(\bx),\ \ \ \psi(\bx)\in \H_0.
\ee
%%%%%%%%%%
In addition, since the Haar measure is invariant, it is in particular invariant under the action of the classical subgroup
$\Theta(\R)$. This action thus naturally gives rise to a one-parameter unitary group ${\cal V} (y)$:
%%%%%%%
\be
\label{newp}
{\cal V}(y)\psi(\bx)= \psi(\bx_y \bx),\ \ \ y\in\R,\ \ \ \psi(\bx)\in \H_0.
\ee

One can easily check that the unitary transformation $\cal T$ maps the operators ${\cal U}(k)$ (\ref{2}) 
to $\Pi(k)$, and $e^{iy{\hat p}}$ (\ref{3}) to ${\cal V}(y)$.
The quantization of the momentum variable can be defined e.g.~on the dense subspace of finite
linear combinations of functions $F_k$, by
%%%
\be
\label{z3}
\Pi(p)F_k=kF_k.
\ee
The $\H_0$ \rep\ defined by (\ref{10}, \ref{newp}) is then a unitarily equivalent version of the polymer \rep.

It is interesting to see how irreducibility is achieved in the configuration \rep, 
despite of the fact that we have only one momentum operator.  This is possible due to the
denseness of the orbits of the action of the classical group $\Theta(\R)$. Thus, there is no
non-trivial function in $\br$ that remains invariant under this action, and therefore no 
configuration operator commutes with $\Pi(p)$.

Finally, note that finite linear combinations of the form $\sum c_k F_k(\bar x)= \sum c_k \bar x(k)$ can be seen as elements of $\H_0$
that depend only on the values of $\bar x$ on a finite number of points. These are called cylindrical functions.
A more general definition will be given in the next section. 
 
\section{Projective aspects}
\label{pro}
The projective characterization of $\br$ is now presented, corresponding to a similar
and very important structure in LQG~\cite{AL1,AL2,AL3,Ba,MM, ALMMT}. As mentioned (see moreover
section~\ref{spect}), this  structure is directly related to the family of finitely
generated $*$-subalgebras of the
configuration $*$-algebra $\cal C$ (\ref{new23}).
To finitely
generated subalgebras correspond finitely generated subgroups of $\R$, which then appear
naturally, playing the role of finitely generated subgroupoids in LQG.
These subgroups, or alternatively  sets of  real numbers  playing the  role of graphs,
are used as a set of labels for both projective and inductive structures.

\subsection{Quantum configuration space as a projective limit}
\label{pro1}
For arbitrary $n\in\N$, a finite set of real numbers $\g=\{k_1,\ldots,k_n\}$ will be said to be independent if 
$k_1,\ldots,k_n$ are algebraically independent with respect to the additive group operation
in $\R$, i.e.\  if the condition
%%%%%%%%%
\be
\label{11}
\sum_{i=1}^n m_i k_i=0, \ \ \ m_i\in\Z,
\ee
%%%%%
implies $m_i=0$, $\forall i$. The set of all such independent sets $\g$ will be denoted by $\Gamma$. 
A partial order relation making  $\Gamma$ a directed set can be introduced as follows.\footnote{Directed
sets can be used as labeling sets for families of objects, generalizing discrete sequences and allowing
consistent definitions of limits.} 
Let $G_\g$ denote the subgroup of $\R$ freely generated by the set $\g=\{k_1,\ldots,k_n\}$:
%%%%%%%%%%%%%%%%
\be
\label{12}
G_\g:=\left\{\sum_{i=1}^n m_i k_i,\ m_i\in\Z\right\}.
\ee
%%%%%%%%%%%%%%%%%%%
A set $\g'$ is said to be greater than $\g$, and we write
$\g'\geq\g$, if $G_\g$ is a subgroup of $G_{\g'}$. It is clear that given $\g$ and $\g'$ one can 
always find $\g''$ such that  $\g''\geq\g$ and 
$\g''\geq\g'$, and so $\Gamma$  becomes a directed set. 
Also, $\Gamma$ has no maximal element.

We are now in position to describe the projective structure of $\br$.
For each $\g\in\Gamma$, let us consider the space (again a group) $\R_\g$ of all morphisms from $G_\g$ to $T$,
%%%%%%%%%%%%%%%
\be
\label{17}
\R_\g:={\rm Hom}[G_\g,T].
\ee
For any pair $\g$, $\g'$ such that $\g'\geq\g$ there are surjective projections
%%%%%%%%%%
\be
\label{18}
p_{\g\g'}: \R_{\g'}\to\R_\g
\ee
defined by restrictions, i.e. an element $\bar x_{\g'}\in\R_{\g'}$ is mapped to its restriction to $G_\g\subset G_{\g'}$.
The projections (\ref{18}) clearly satisfy the consistency conditions
%%%%%%%%%
\be
\label{19}
p_{\g\g''}=p_{\g\g'}p_{\g'\g''}, \ \ \forall \g''\geq\g'\geq\g.
\ee
%%%%%%%%%%%%%%%%%%%%
Such a family of spaces and projections, labeled by a directed set, is called a projective
family.  

Since each $G_\g$ is freely generated by an independent set $\g=\{k_1,\ldots,k_n\}$, each $\R_\g$ is essentially
a space of the form $T^n$, where $n$ is the cardinality of the corresponding set $\g$. In fact, every element
$\bar x_\g$ of $\R_\g$ is uniquely determined by the images of the generators, and  one gets a bijection
between $\R_\g$ and $T^n$, given by $\bar x_\g\mapsto \left( \bar x_\g(k_1),\ldots,\bar x_\g(k_n)\right)$.  
From now on we identify each $\R_\g$  with the corresponding space $T^n$ with the standard topology, 
or with the $n$-torus
$\R^n/(2\pi\Z)^n$. (We will not distinguish between the two, 
i.e.~the correspondence $\R/(2\pi\Z)\ni x\leftrightarrow e^{ix}\in \C$ is freely used.)
%where $U(1)$ is identified with $\left(\R/(2\pi)\right)$.
%\times\cdots\times\left(\R/(2\pi/k_n)\right)$.
Each space $\R_\g$ is then a compact
space. It can be seen that the projections (\ref{18}) are continuous,
and so the family $\{\R_\g\}_{\g\in\Gamma}$ forms a  compact  projective family.

There is a well defined notion of projective limit of a family of spaces, 
which in the case of a compact family  is again a compact space.
The projective limit of the family $\{\R_\g\}_{\g\in\Gamma}$ is the subset of the cartesian product $\times_{\g\in\Gamma}\R_\g$
of those elements $(\bar x_\g)_{\g\in\Gamma}$ that satisfy the consistency conditions
%%%%%%%%%%%%
\be
\label{20}
p_{\g\g'}\bar x_{\g'}=\bar x_\g, \ \ \forall \g'\geq\g.
\ee
Not surprisingly,  there is a bijection between the projective limit of the family $\{\R_\g\}_{\g\in\Gamma}$ and
the set $\br={\rm Hom}[\R,T]$, given by
%%%%%%%%%%%%
\be
\label{21}
\br\ni \bar x \mapsto (\bar x_{|\g})_{\g\in\Gamma},
\ee
where $\bar x_{|\g}$ denotes the restriction of $\bar x$ to the subgroup $G_\g$.
Moreover, the natural topology on the projective limit, namely the weakest topology
such that all projections 
%%%%%%%%%%%%
\be
\label{22}
p_\g:\br\to\R_\g,\ \ \bar x \mapsto \bar x_{|\g},
\ee
%%%%%
are continuous,
coincides with the Tychonoff topology introduced in  section \ref{group}.

Thus, the compact space $\br$ is naturally homeomorphic to the projective limit
of the family $\{\R_\g\}_{\g\in\Gamma}$. Since each space $\R_\g$ is in turn
homeomorphic to a torus, one can see the 
space $\br$ as a limit of a familiy of tori, of 
growing dimensions.

Although the projective description of the quantum configuration space 
is analogous to the one encountered
in LQG [with tori $T^n$  corresponding to manifolds $SU(2)^n$], 
there is a difference worth mentioning, related to the fact that
we now have a 1-dimensional classical configuration space $\R$, while most of the
spaces $\R_\g$ are multi-dimensional. Thus, the projection $p_\g\Theta(\R)$ to a generic $\R_\g$
(e.g.\ with $\g$ containing at least two elements) of the image
 $\Theta(\R)\subset\br$ does not cover $\R_\g$. The image of $\R$ on $\R_\g$
is nevertheless dense. In fact, writing $\g=\{k_1,\ldots,k_n\}$ and using the identification of  $\R_\g$ with
the $T^n$ torus $\R^n/(2\pi\Z)^n$, one can see that $\R$ is mapped to $T^n$ as follows:
%%%%
\be
\label{toro}
\R\ni x\mapsto (k_1x,\ldots,k_nx)\in T^n. 
\ee
We therefore obtain an injective map $p_\g\Theta:\R\to T^n\cong \R_\g$
with dense image, since the frequencies $\{k_1,\ldots,k_n\}$ are algebraically independent.

\subsection{Cylindrical functions and measure theoretical aspects}
\label{irrel}

We will now consider the pull-back of the projections (\ref{22}) above. For any given $\g$ and any given
continuous function $f\in C(\R_\g)$, the pull-back $p_\g^*f(\bar x)=f(p_\g\bar x)$ gives a function in $\br$.
Since the projections are continuous and surjective, the function $p_\g^*f$ is again continuous
and the map $p_\g^* : C(\R_\g)\to C(\br)$ is injective. Thus, functions on any given $\R_\g$, or
corresponding tori, are faithfully mapped to functions in $\br$. These so-called cylindrical functions
are therefore functions in $\br$ that are essentially living on a torus. Moreover, there is a "sufficient"
supply of these functions, in the sense that the set of all such functions, for all $\g\in\Gamma$, is dense in 
$C(\br)$, and therefore $L^2$-dense. Of course, one cannot stop at any
fixed $\g$, but note that {\it finite} linear combinations of cylindrical functions corresponding to 
different $\g$'s, say $\g_1,\ldots,\g_n$, can again be seen as functions on a higher dimensional
torus, since there is $\g'$ such that $\g'\geq\g_1,\ldots,\g'\geq\g_n$.

Furthermore, the pull-back extends from a map between continuous functions to a map between $L^2$-spaces.
To see this and for subsequent discussion, let us introduce a simple notion from measure theory.

Since $\br$ projects to every space $\R_\g$, the Haar measure $\m_0$ defines a measure ${\m_0}_\g$ 
on each $\R_\g$. Explicitly, each measure  ${\m_0}_\g$  is defined by push-forward of $\m_0$, with 
respect to the projection $p_\g:\br\to\R_\g$, i.e.
%%%%%
\be
\label{compm}
{\m_0}_\g(B)=\m_0(p_\g^{-1}B)\ \ \forall\ {\rm measurable\ set}\ B\subset \R_\g,
\ee
where  $p_\g^{-1}B$ denotes the inverse image of the set $B$. An equivalent definition of
${\m_0}_\g$ can be given in terms of integration of continuous functions, namely ${\m_0}_\g$
is defined by
%%%%%%%%%
\be
\label{32}
\int_{\R_\g}f d{\m_0}_\g=\int_{\br}p^*_{\g}f d\m_0, \ \ \ \forall f\in C(\R_\g).
\ee
%%%%%%%%
As expected,  measures ${\m_0}_\g$ coincide with Haar measures on the groups $\R_\g$, or the 
usual (Haar - Lebesgue) measures on  corresponding 
tori.\footnote{The Ashtekar-Lewandowski measure in LQG was actually introduced the other way around,
using a (suitably compatible) family of Haar measures on finite dimensional spaces of a projective family
to define a measure on the projective limit $\ab$. This is the general procedure to introduce measures in projective
limit spaces with no further structure. In amenable cases, including linear field theories and LQG, expressions
like (\ref{compm}), seen as the definition of the measure on the projective limit, indeed lead to
proper $\sigma$-aditive measures. In the  case of compact families, the formalism of $C^*$-algebras representation theory, corresponding to (\ref{32}),
is also available. (See~\cite{AL1,AL2,MM} for measure theoretical aspects in LQG.)}
(This can be easily checked e.g.~from (\ref{32}).)

One can then form the Hilbert spaces $L^2(\R_\g,{\m_0}_\g)$, which are essentially the usual
Hilbert spaces of square integrable functions on tori. It is easily seen (again from (\ref{32})) that,
for each $\g$, the pull-back defines a unitary transformation between  $L^2(\R_\g,{\m_0}_\g)$
and the separable Hilbert subspace ${\H_0}_\g:=p_\g^*L^2(\R_\g,{\m_0}_\g)$ of $\H_0$.

In particular, one can consider the restriction of the polymer \rep\ to a given  ${\H_0}_\g$,
with $\g=\{k_1,\ldots,k_n\}$. Of course, operators $\Pi(k)$ (\ref{10}) with $k$ outside the group $G_\g$
no longer act on ${\H_0}_\g$, but those who do, i.e.~such that $k=\sum m_j k_j$,
are seen to correspond to the usual quantization
of the functions $\exp (i\sum m_j x_j)$ on the torus $T^n$. 
On the other hand, the momentum operator $\Pi(p)$ (\ref{z3}) is defined on every ${\H_0}_\g$,
and corresponds to derivation along the direction of the torus defined by $p_\g\Theta(\R)$ as above.

Finally, let us  see that  the image $\Theta(\R)$ of the classical configuration space 
is a zero Haar measure set in $\br$. 
This is particularly easy to see in the present case, since it already happens for generic $\R_\g$.
Consider then a set $\g=\{k_1,k_2\}$ with two elements (or more), so that the space $\R_\g$
can be seen as  $T^2$ (or $T^n$, $n\geq 2$.)
Since the number of times that the line $\R$ is wrapped around the torus is clearly countable,
the measure of the set $p_\g\Theta(\R)$ is zero. (The situation is similar to that of the set
of lines of constant rational $y$ in the square $(x,y)\in[0,1]\times[0,1]$.) 	
It follows immediately from (\ref{compm}) that $\Theta(\R)$ is a zero Haar measure set in $\br$.

As a consequence, the restriction of a generic element of $\H_0=L^2(\br,\m_0)$ to $\Theta(\R)$
is not defined. 
%%%
%It is true  that it is defined for continuous functions,  but $\C(\br)$ is not complete
%in the $L^2$-norm. 
In this sense, the completion of the linear space $\cal C$ (\ref{new23})
with respect to the inner product $\langle e^{ikx}, e^{ik'x}\rangle=\delta_{k k'}$ leads
to a space of functions in $\br$,  not on $\R$.\footnote{In other words, the non-continuous function
$\R\ni k\mapsto \delta_{k 0}$ is not the Fourier transform of a measure in $\R$, but in $\br$.
On the momentum side, the discrete measure is defined on $\R$, but it is not $\sigma$-finite,
allowing it to produce a unitary implementation of translations 
and be non-equivalent to the Lebesgue measure.} Thus, the extension of the configuration space
from
$\R$ to $\br$ is essential in the polymer \rep.

%%%%%%%%%%%%%%%%%%%%%%%%%%%%%%
%%%%%%%%%%%%%%%%%%%%%%%%%%%
\section{$C^*$-algebra aspects}
\label{spect}
We will now see explicitly that $\br$ is in fact the spectrum of the $C^*$-algebra of almost periodic 
functions in $\R$~\cite{R}. This characterization of the quantum configuration space corresponds to the
original introduction of the space of generalized connections $\ab$ as the spectrum of the holonomy 
algebra~\cite{AI,AL1}.

Let us  consider again the configuration $*$-algebra $\cal C$ of  functions  given by finite sums of
the form
%%%%%%%%%%
\be
\label{23}
\R\ni x\mapsto f(x)=\sum_j c_j e^{ik_jx},
\ee
and its $C^*$-completion $\bar{\cal C}$, with respect to the supremum norm,
$\|f(x)\|={\rm sup}_{x\in\R}|f(x)|$. 

The spectrum $\Delta(\bar{\cal C})$ of the algebra $\bar{\cal C}$
is the set of all non-zero multiplicative linear functionals on $\bar{\cal C}$, i.e.~non-zero linear functionals
$\vf: \bar{\cal C}\to\C$
such that
%%%%%%%%%
\be
\label{25}
\vf(fg)=\vf(f)\vf(g),\ \ \forall f,g\in \bar{\cal C}.
\ee
%%%%%%%
(Such functionals are necessarily continuous.) One  can easily check that 
$\vf(e^{ikx})$ takes values in $T$, $\forall \vf\in \Delta(\bar{\cal C})$, $\forall k$,
and it follows that the
assignment $\R\ni k\mapsto  \vf(e^{ikx})$ determines a group morphism, i.e.\ an element 
of $\br$. There is thus a map,  injective, from $\Delta(\bar{\cal C})$ to $\br$.
Conversely,  any element $\bar x\in\br$ 
defines a non-zero multiplicative linear functional on $\cal C$, by $e^{ikx}\mapsto \bar x(k)$.

To establish a bijection between  $\Delta(\bar{\cal C})$ and $\br$ it
remains to prove that these functionals can be extended to functionals on $\bar{\cal C}$,
for which it is sufficient to show that they are continuous on $\cal C$. This is another 
point were one finds a difference to a corresponding argument in LQG, again due to the above mentioned fact
that $p_\g\Theta(\R)$ does not cover $\R_\g$, for generic $\g$. However, the denseness of $p_\g\Theta(\R)$
leads to the same result, 
as we now confirm.\footnote{Similar  arguments nevertheless appear in Appendix A of~\cite{AL1},
where $U(1)$ holonomies associated with piecewise $C^1$ (rather than piecewise analytic) loops
are considered.}

We begin by displaying the inductive structure of the algebra ${\cal C}$.
For each $\g\in\Gamma$, $\g=\{k_1,\ldots,k_n\}$, let ${\cal C}_\g\subset {\cal C}$ denote the $*$-subalgebra
generated by the set of functions $\{e^{ik_1x},\ldots,e^{ik_nx}\}$, whose elements
are finite sums of the form
%%%%%%
\be
\label{cg}
f(x)=\sum_k c_k e^{ikx}  \ \ \ {\rm with}\ \ \ k\in G_\g.
\ee
%%%%
It is clear that  ${\cal C}_\g$ is a subalgebra of ${\cal C}_{\g'}$, whenever
$\g'\geq \g$. Moreover, any element of ${\cal C}$ belongs to some subalgebra ${\cal C}_\g$.
In fact, since the number  of different frequencies $k_j$ in (\ref{23}) is
finite $\forall f\in \cal C$, they all belong to some group $G_\g$. Thus, the algebra ${\cal C}$ is the union of all the
subalgebras ${\cal C}_\g$.

Let then $f(x)=\sum_j c_j e^{ik_jx}$ be an arbitrary element of $\cal C$, and 
$\g=\{\l_1,\ldots,\l_n\}$ an independent set such that $k_j\in G_\g$ $\forall j$, 
i.e.
%%%%%%%%
\be
\label{26}
k_j=\sum_{l=1}^n m^j_l\l_l,\ \ \ \forall j,
\ee
for some integers $m^j_l$.
We thus have
%%%%%%%%%%%%
\be
\label{27}
\|f(x)\|={\rm sup}_{x\in\R}\left|\sum_j c_j\prod_{l=1}^n(e^{i\l_lx})^{m^j_l}\right|.
\ee
%%%%%%%%%%%%%%%
For each value of $x$, $(e^{i\l_1x},\ldots,e^{i\l_nx})$ is an element of 
$T^n\cong \R_\g$, and the image of $\R$ by this map 
is a dense set.
Since functions in $T^n$ of the form 
%%%
\be
\label{u1}
T^n\ni(x_1,\ldots,x_n)\mapsto \sum_j c_j\prod_l (e^{ix_l})^{m^j_l}
\ee
(with finite sums) are continuous, one can see that the supremum over $\R$ in 
(\ref{27}) can be replaced by the supremum over $T^n$. Thus
%%%%%%%%%%%%
\be
\label{28}
\|f(x)\|={\rm sup}_{(x_1,\ldots,x_n)\in T^n}\left|\sum_j c_j\prod_{l=1}^n(e^{ix_l})^{m^j_l}\right|.
\ee
It is now easy to see that linear functionals on $\cal C$ defined by elements of $\br$
are continuous. Let then   $\vf_{\bar x}$ denote the functional defined
by $\vf_{\bar x}(e^{ikx})=\bar x(k)$, $\bar x\in\br$. We obtain
%%%%%%%%
\be
\label{29}
\vf_{\bar x}\left(f(x)\right)=\sum_j c_j \bar x(k_j)=\sum_j c_j\prod_{l=1}^n\bar x(\l_l)^{m^j_l},
\ee
which is simply the evaluation of the function appearing in (\ref{28}),
at the 
%particular  
point
$(\bar x(\l_1),\ldots,\bar x(\l_n))\in T^n$.
Thus, $|\vf_{\bar x}\left(f(x)\right)|\leq \|f(x)\|$, $\forall f(x)\in\cal C$, proving that the functionals
$\vf_{\bar x}$ are continuous.

A bijection between the spectrum
$\Delta(\bar {\cal C})$ and $\br$ is therefore established.
Moreover,   $\Delta(\bar {\cal C})$ and $\br$ 
are  homeomorphic. In fact, the topology  on the spectrum
$\Delta(\bar {\cal C})$, namely the  Gelfand topology defined as the weakest one such that
all maps
%%%%%%%%
\be
\label{31}
\Delta(\bar {\cal C})\to\C,\ \ \ \vf\mapsto\vf(f), \ \ \ f\in \bar {\cal C},
\ee
are continuous, is again seen to correspond  to the Tychonoff topology  in $\br$ 
introduced in section \ref{group}. It is also clear that the Gelfand compactification $\R\to \Delta(\bar {\cal C})$,
given by $y\mapsto \vf_y$ such that $\vf_y(e^{ikx})=e^{iky}$, corresponds to the map $\Theta$ (\ref{theta}).

Finally, let us make  the relation between inductive structures and projective
structures more explicit.

Considering  the completions ${\bar {\cal C}}_\g$ of the algebras ${\cal C}_\g$,
it is again true that ${\bar{\cal C}}_\g\subset {\bar{\cal C}}_{\g'}$ for $\g'\geq \g$, but the 
union $\cup_\g {\bar{\cal C}}_\g$ is not a complete space.
The family $\{{\bar{\cal C}}_\g\}_{\g\in\Gamma}$, with natural inclusions, is a particular
case of an inductive family of $C^*$-algebras. 
The completion of the union,  in this case $\bar {\cal C}$, is called in this context the inductive limit $C^*$-algebra.
%It coincides in  this case  with $\bar {\cal C}$, of course.
Taking now into account 
that the set of all functions
of the form (\ref{u1}) is  dense  in $C\left(T^n\right)$, the above arguments
immediately lead to the conclusion that each algebra
${\bar {\cal C}}_\g$ is isomorphic to the algebra $C(\R_\g)$.
In other words,  $\R_\g$ is the spectrum of ${\bar {\cal C}}_\g$.
As we take the limit over the set $\Gamma$, we find that $\br$ is the spectrum 
of ${\bar {\cal C}}$.\footnote{There is a perfect duality between compact projective families and
inductive families of $C^*$-algebras with identity, in the sense that there is a one-to-one
(Gelfand) correspondence and the compact projective limit is the spectrum of the
corresponding inductive limit $C^*$-algebra (see~\cite{AL2}).}

Summarizing, from the  point of view of $C^*$-algebras, the configuration $C^*$-algebra ${\bar {\cal C}}$
and the algebra $C(\br)$ of continuous functions in $\br$ are one and the same.
As in the finite dimensional cases seen above, this can be understood from the 
denseness of $\Theta (\R)$, allowing for the restriction
of continuous functions to  be an isomorphism.

%%%%%%%%%%%%%%%  Acknowledgments %%%%%%%%%%%%%%%%%%%%%%

\subsubsection*{Acknowledgements}
\noindent 
I am greatly thankful to Jos\'e Mour\~ao, Guillermo Mena Marug\'an, Luis Garay and 
Mercedes Mart\'\i n-Benito.
This work was supported in part by 
%projects 
%POCTI/33943/MAT/2000, CERN/P/FIS/43171/2001 and POCTI/FNU/\-49529/2002.
POCTI/FIS/57547/2004.
%%%%%%%%%%%%%%%%%%%%%%%%%%%%%%%%

\section*{Appendix: The Schr\"{o}dinger representation}
\label{append}
We will now see how the usual Schr\"{o}dinger representation is obtained in the present
context. It is given, of course, by a different measure in $\br$.
Though technically much simpler, the introduction of this measure corresponds to the construction
of the so called $r$-Fock measures in the loop quantization of 
$U(1)$ connections~\cite{Va,Vc,ALfock} (see also~\cite{ALS}).

Let us start by considering the Gaussian measure $d\nu_G=e^{-x^2/2}{dx\over\sqrt{2\pi}}$ in $\R$, and the 
corresponding space of square integrable functions $L^2(\R,\nu_G)$.
The Hilbert space $L^2(\R,\nu_G)$ carries (a \rep\ unitarily equivalent to) the usual Schr\"{o}dinger
\rep\
of the Weyl relations.

We will now use the 
natural dense injection $\Theta:\R\to\br$ (\ref{theta}) in order to  push-forward the Gaussian measure,
thus obtaining a measure
$\m_G$ in $\br$. The measure
$\m_G$ is defined by
%%%%%
\be
\label{41}
\m_G(B)=\nu_G(\Theta^{-1}B)\ \ \forall\ {\rm measurable\ set}\ B\subset \br.
\ee
The corresponding Hilbert space $L^2(\br,\m_G)$  is then
isomorphic to  $L^2(\R,\nu_G)$. To confirm it, let us see that 
the mapping between dense subspaces defined
by
%%%
\be
\label{40}
L^2(\R,\nu_G)\ni e^{ikx}\mapsto F_k(\bar x), \ \ k\in\R,
\ee
preserves the inner product, thus defining a
unitary transformation. With $\Theta^*$ denoting pull-back, we  obtain
%%%%%%
\be
\label{42}
\int_{\br}F_k^* F_{k'} d\m_G  = \int_{\br} F_{(k'-k)} d\m_G = \int_{\R} \Theta^* F_{(k'-k)} d\nu_G,
\ee
where  the second equality follows from the definition (\ref{41}) of $\m_G$.
Recalling  definitions   (\ref{7}) and (\ref{theta}), it is clear that $\Theta^*F_k$ is the exponential
function $e^{ikx}$, and therefore
%%%
\be
\label{43}
\int_{\br}F_k^* F_{k'} d\m_G =  \int_{\R} e^{i(k'-k)x} d\nu_G.
\ee

The measure $\m_G$ provides a \rep\ of the configuration algebra which is obviously different from the polymer \rep.
The functions $x\mapsto e^{ikx}$ are again quantized as the multiplication
operators $F_k$, as in (\ref{10}), with the crucial difference that the measure
$\m_G$ is now supported on the image $\Theta (\R)$. Thus, essentially, only points $x\in\R$ of the
classical configuration space contribute to the measure, the functions $F_k$ reassume the form $e^{ikx}$,
and continuity with respect to $k$ is restored.  It is clear that the measure $\m_G$ gives a representation
of the Weyl relations which is equivalent to the Schr\"{o}dinger one, the unitary map being transformation (\ref{40}).
The Haar measure $\m_0$ and the measure $\m_G$ are mutually singular,
since $\Theta (\R)$ is a set of zero Haar measure.

%%%%%%%%%%%%%% Bibliography %%%%%%%%%%%%%%%%%%%%%%%

%%%%%%%%%%%%%%%%%%%%%%%%%%%%%%%%%%

%%%%%%%%%%%%%%%%%%%%%%%%%%%%%%%%%

\end{document}